\documentclass[aps,pre,twocolumn,groupedaddress,longbibliography]{revtex4-1}

\usepackage{graphicx}
\usepackage{bm}
\usepackage{amsmath}

\begin{document}

\title{Annihilation trajectory of defects in smectic-C films}
\author{Xingzhou Tang}
\author{Jonathan V. Selinger}
\affiliation{Department of Physics, Advanced Materials and Liquid Crystal Institute, Kent State University, Kent, Ohio 44242, USA}

\date{June 4, 2020}

\begin{abstract}
In a 2D liquid crystal, each topological defect has a topological charge and a characteristic orientation, and hence can be regarded as an oriented particle. Theories predict that the trajectories of annihilating defects depend on their relative orientation. Recently, these predictions have been tested in experiments on smectic-C films. Those experiments find curved trajectories that are similar to the predictions, but the detailed relationship between the defect orientations and the far-field director is different. To understand this difference, we extend the previous theories by adding the effects of elastic anisotropy, and find that it significantly changes the curved trajectories.
\end{abstract}

\maketitle

\section{Introduction}

The dynamics of topological defects is an important feature of the physics of liquid crystals. In many experiments, defects form when a disordered phase is quenched into a more ordered phase, such as an isotropic into a nematic phase, and then the material coarsens as defects annihilate each other~\cite{Chuang1991,Pargellis1991,Bowick1994,Oswald2005,Blanc2005,Dierking2012,Guimaraes2013,Kim2013}. A particularly simple case is provided by thin films of smectic liquid crystals. In a smectic-C phase, the director tilt is a vector order parameter, and it allows the formation of defects with topological charges of $\pm1$. Defects of opposite topological charge attract each other and annihilate, and this dynamic process can be observed through optical microscopy~\cite{Muzny1992,Pargellis1992,Stannarius2006,Stannarius2016}.

For many years, there is been a standard, textbook-style theory of defect annihilation in smectic-C films~\cite{Kleman2003}. In this theory, defects of opposite topological charge $+1$ and $-1$ attract each other through an potential that scales logarithmically with defect separation $r$, and hence a force that scales as $1/r$, analogous to Coulomb's law in two dimensions (2D)~\cite{Dafermos1970}. This elastic force competes with a drag force, which is proportional to the defect velocity, with a coefficient that is the same for the $\pm1$ defects~\cite{Imura1973}. Based on the balance of elastic and drag forces, this theory predicts that the defects move straight together, the two defects have the same speed as each other, and this speed increases until they annihilate.

Theoretical research has investigated two types of deviations from this standard theory. First, studies have considered corrections due to \emph{backflow}, i.e.\ fluid flow induced by liquid crystal director rotation. This effect was explored first through numerical calculations~\cite{Toth2002,Svensek2003}, and further through approximate analytic arguments~\cite{Kats2002,Sonnet2005,Sonnet2009,Tang2019}. These studies find that backflow has very different effects on different topological charges: It tends to enhance the motion of positive defects but inhibit the motion of negative defects. Hence, during the defect annihilation process, the $+1$ defect moves more rapidly than the $-1$ defect. As a result, the annihilation event occurs closer to the initial position of the $-1$ defect, and farther from the initial position of the $+1$ defect. Some of these studies report that a similar effect can be caused by elastic anisotropy, i.e.\ difference of Frank elastic constants.

A second deviation from the standard theory is related to \emph{defect orientation}. Recent theoretical studies of have emphasized that each defect has an orientation in the 2D plane. Vromans and Giomi have characterized the orientation by a vector~\cite{Vromans2016}, and we have proposed an alternative description based on tensors~\cite{Tang2017}. Defect orientation is especially important for active nematic liquid crystals~\cite{Shankar2018}, but the concept also applies to conventional liquid crystals, and to phases other than nematic. For smectic-C films, the $+1$ defect is a special case because it does not have an orientation, but instead has a scalar property indicating whether it has splay, bend, or a mixture. By comparison, the $-1$ defect has an orientation with two-fold symmetry in the plane. The energy of two interacting defects depends on their relative orientation as well as on their separation; the classic prediction of a logarithmic potential applies only in the case of the optimal relative orientation. For a non-optimal relative orientation, the interaction is not a central-force potential; it does not attract two defects straight together. (This form of the potential is implicit in Ref.~\cite{Tang2017}, and is discussed explicitly in Ref.~\cite{Missaoui2020} and in Sec.~IV below.) As a result, the defects are predicted to follow an S-shaped trajectory as they move together and annihilate~\cite{Vromans2016,Tang2017}.

In a recent experimental study~\cite{Missaoui2020}, Missaoui \emph{et al.}\ observed the annihilation of $\pm1$ defects in smectic-C films, and compared the observed trajectories with the predictions of Ref.~\cite{Tang2017}. They found S-shaped trajectories that were qualitatively similar to the predicted trajectories. However, the detailed relationship between the defect orientations and the far-field director was different from the prediction, assuming that the system evolves through a series of quasiequilibrium director configurations as the defects approach each other. They suggested that the time evolution might be more complex, so that the system does not evolve through such quasiequilibrium states.

The purpose of this article is to investigate the effects of unequal Frank elastic constants on defect trajectories. In general, smectic-C films have different elastic constants for splay and bend deformations. The difference between elastic constants is already known to affect the structure of defects in a 2D liquid crystal~\cite{Ranganath1983,Hudson1989,Zhang2018}, as well as the relative speed of positive and negative topological charges~\cite{Toth2002,Svensek2003}. Here, we show that it can also create S-shaped annihilation trajectories, which is the same effect as a non-optimal relative orientation of the defects. Hence, unequal Frank constants must be taken into account when analyzing experimental defect trajectories, and this mechanism may provide an alternative explanation for the experimental results of Ref.~\cite{Missaoui2020}.

The plan of this paper is as follows. In Sec.~II, we present the simulation method for relaxational dynamics of a smectic-C film without fluid flow. In Sec.~III, we apply this method to the motion of a single defect. In Sec.~IV, we extend it to calculate the trajectories of two oppositely charged defects during the annihilation process. In Sec.~V, we discuss the conclusions of this study.

\section{Simulation method}

We consider a smectic-C liquid crystal in a 2D film geometry, with no activity. This phase has tilt order, which can be described by the vector field $\textbf{\textit{P}}(\textbf{\textit{r}},t)$, representing the projection of the liquid crystal director field into the 2D plane. This vector field may have disclinations, which are point defects in the plane, with integer topological charges. In our previous articles, we discussed three ways to understand the dynamics of such defects. The first and most standard method is to solve partial differential equations to describe the time evolution of the orientational order everywhere in the material. These partial differential equations may represent relaxational dynamics without fluid flow~\cite{Tang2017}, or hydrodynamics that couples changes in orientational order with fluid flow~\cite{Tang2019}. A second method is to regard topological defects as oriented particles with effective interactions and drag coefficients, and solve the dynamics macroscopically~\cite{Tang2019}. A third approach is to consider the effects of shear flow as an effective potential acting on the orientational order~\cite{Tang2020}. In this work, we take the first approach, and use pure relaxational dynamics so that we can see the effects of unequal Frank constants without backflow.

To describe a system in which the defects are free to move, we use an order parameter with variable magnitude and direction,
\begin{equation}
\bm{P}(\bm{r},t)=P(\bm{r},t)
\begin{pmatrix}
\cos\theta(\bm{r},t) \\
\sin\theta(\bm{r},t)
\end{pmatrix}.
\end{equation}
In terms of this order parameter, the free energy can be expressed as
\begin{align}
F=\int d^2 r \biggl[&-\frac{1}{2}a|\bm{P}|^2+\frac{1}{4}b|\bm{P}|^4+\frac{1}{2}k_S(\bm{\nabla}\cdot\bm{P})^2\nonumber\\
&+\frac{1}{2}k_B(\bm{\nabla}\times\bm{P})^2 \biggr].
\end{align}
Here, the first two terms are an expansion of the free energy in powers of $P$, and they favor the magnitude $P_\text{bulk}=(a/b)^{1/2}$ in the bulk, where gradients are small. The last two terms are the Frank elastic free energy for splay and bend, respectively. To model the time evolution, we use the equations for relaxational dynamics
\begin{align}
\frac{\partial P_x(\bm{r},t)}{\partial t}=-\frac{1}{\gamma_1}\frac{\delta F}{\delta P_x(\bm{r},t)},\nonumber \\
\frac{\partial P_y(\bm{r},t)}{\partial t}=-\frac{1}{\gamma_1}\frac{\delta F}{\delta P_y(\bm{r},t)},
\label{dynamicequations}
\end{align}
where $\gamma_1$ is the rotational viscosity.

\section{Single defect with unequal Frank constants}

\begin{figure}
\includegraphics[width=\columnwidth]{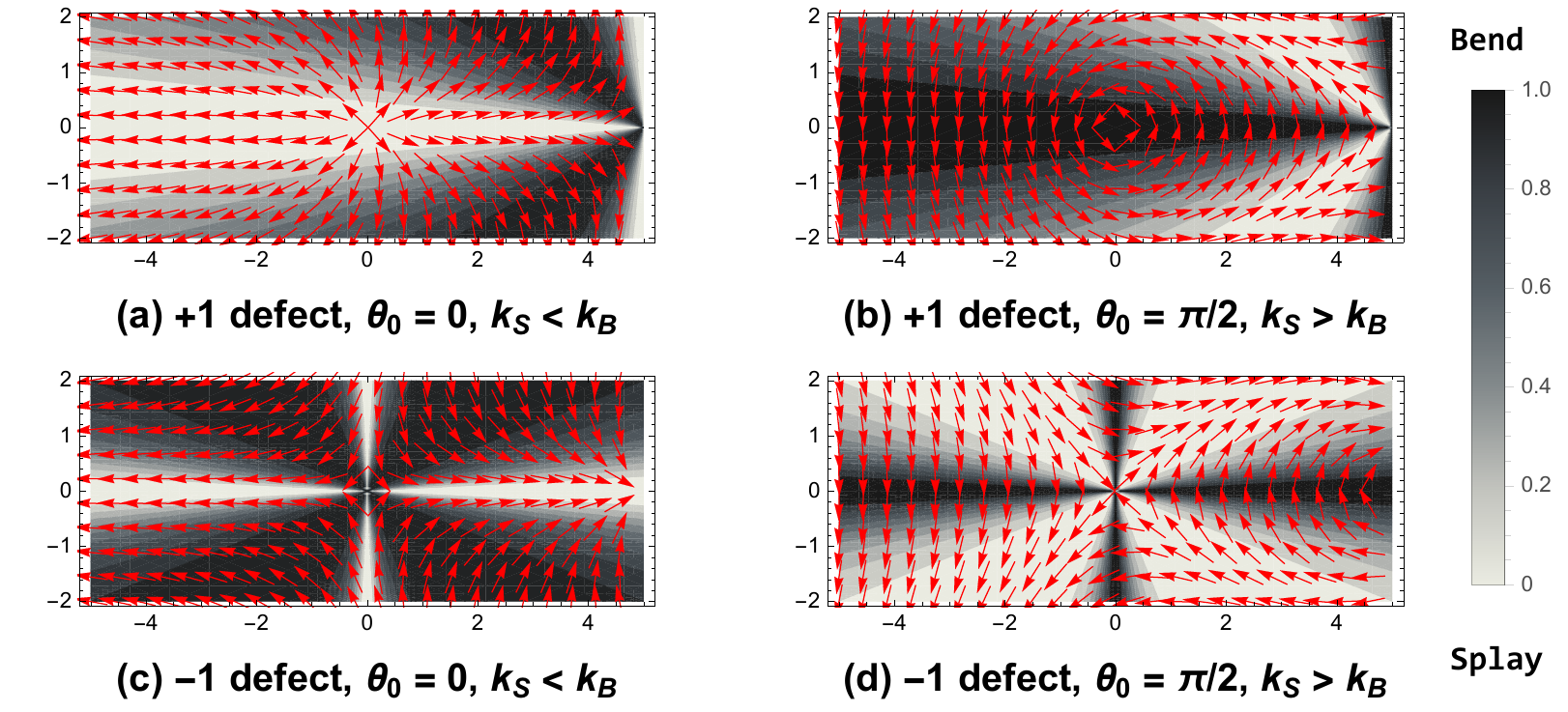}
\caption{Structure of $\pm1$ defects with different values of $\theta_0$. The black arrows indicate the initial conditions for the director field, before the defects begin to move. The background colors indicate the local ratio $F_B/(F_B+F_S)$. White areas are mostly splay, while black areas are mostly bend.}
\end{figure}

In this section, we focus on the dynamics of a single defect in a smectic-C liquid crystal. First, consider a defect of topological charge $+1$. Around the defect, the direction of orientational order is
\begin{equation}
\theta(x,y,t)=\tan^{-1}\left[\frac{y-y_0(t)}{x-x_0(t)}\right]+\theta_0(t).
\end{equation} 
In this equation, $(x_0(t),y_0(t))$ is the location of the defect at time $t$, and $\theta_0(t)$ is an angle that describes the structure of the defect. If $\theta_0=0$, the orientational order points radially outward from the defect and the deformation is pure splay, as in Fig.~1(a). If $\theta_0=\pi$, the director field points radially inward toward the defect and the deformation is again pure splay. If $\theta_0=\pm\pi/2$, the director points tangentially around the defect and the deformation is pure bend, as in Fig.~1(b). If $\theta_0$ has an intermediate value, then the deformation has a combination of splay and bend. In a liquid crystal with equal Frank constants for splay and bend, these structures are all equal in energy. More generally, in a liquid crystal with unequal Frank constants, these structures have different energies. Hence, in this general case, each $+1$ defect is driven toward a specific value of $\theta_0$, corresponding to either pure splay or pure bend, whichever has a lower Frank constant.

The considerations are quite different for a defect of topological charge $-1$. If the Frank constants are equal, the direction of orientational order around the defect is
\begin{equation}
\theta(x,y,t)=-\tan^{-1}\left[\frac{y-y_0(t)}{x-x_0(t)}\right]+\theta_0(t).
\label{thetaminus1}
\end{equation} 
Following the argument of Ref.~\cite{Tang2017}, the director points radially outward at an angle of $\psi=\theta_0/2$ (mod $\pi$), and radially inward at an angle of $\psi\pm\pi/2$. The angle $\psi$ can be regarded as the orientation of the $-1$ defect. Changing $\theta_0$ does not change the structure of the defect, but rather changes the orientation. For example, in Figs.~1(c,d), the parameter $\theta_0$ rotates by $\pi/2$, and hence the orientation $\psi$ rotates by $\pi/4$. If the Frank constants are unequal, then the director field is more complex than Eq.~(\ref{thetaminus1}). Even so, the defect still has two-fold symmetry, and the director field still points radially outward in two directions, which can be labeled as $\psi$ (mod $\pi$). Hence, the angle $\theta_0=2\psi$ (mod $2\pi$) can still be defined. Regardless of $\theta_0$, the $-1$ defect has a mixture of splay and bend. Because $\theta_0$ does not affect the relative amounts of splay and bend, $\theta_0$ is not driven toward a specific value, but rather is free to vary.

To study the influence of unequal Frank constants on the motion of a single defect, we construct a model in which a test defect is attracted toward an opposite-charged defect fixed on the boundary. Figures~1(a,b) show initial conditions for a test defect of charge $+1$ at the center, with a $-1$ defect is fixed on the right boundary. Figures~1(c,d) show initial conditions for a test defect of charge $-1$ at the center, with a $+1$ defect is fixed on the right boundary. In addition to the initial configurations of the director field, the figures also show a color indicating the local ratio $F_B/(F_B+F_S)$, where $F_S=\frac{1}{2}k_S(\bm{\nabla}\cdot\bm{P})^2$ and $F_B=\frac{1}{2}k_B(\bm{\nabla}\times\bm{P})^2$ are the local free energy densities of splay and bend, respectively. Here, white regions are mostly splay, and black regions are mostly bend. These figures are based on analogous figures in Ref.~\cite{Missaoui2020}.

\begin{figure}
\includegraphics[width=\columnwidth]{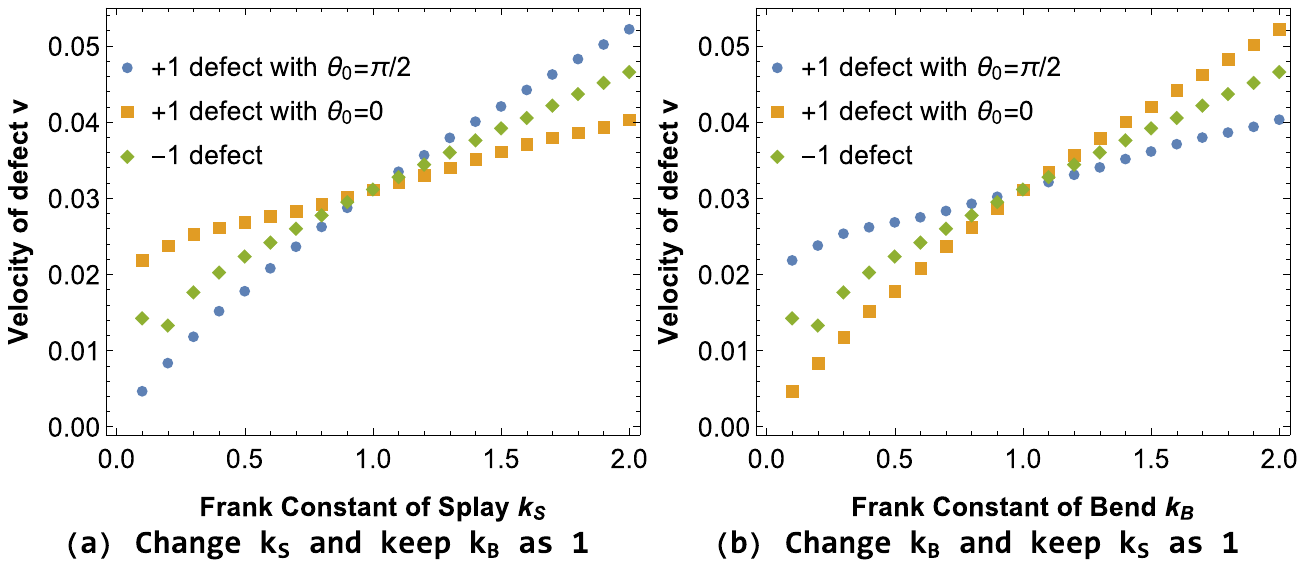}
\caption{Velocity of defects with different topological charges and different $\theta_0$ angles as functions of the Frank constants. (a) Varying $k_S$ with fixed $k_B=1$. (a) Varying $k_B$ with fixed $k_S=1$.}
\end{figure}

We now allow the field $\bm{P}(\bm{r},t)$ to evolve in time, following the dynamic equations~(\ref{dynamicequations}). As it evolves, the test defect moves toward the opposite defect fixed on the boundary, and we measure the velocity. Figure~2 shows the velocities of defects with different topological charges and different $\theta_0$ angles as functions of the Frank elastic constants. In Fig.~2(a), $k_B$ is fixed at 1 and $k_S$ is changed from 0.1 to 2, while in Fig.~2(b), $k_S$ is fixed at 1 and $k_B$ is changed from 0.1 to 2. In both plots, the general trend is that the defect velocity increases as the Frank constants increase. However, the details depend on the type of defect. For a $-1$ defect, the velocity does not depend on the initial $\theta_0$, and the velocity follows the same trend regardless of which Frank constant varies. By contrast, for a $+1$ defect, the velocity is sensitive to the initial $\theta_0$, and the trend depends on which Frank constant varies. This dependence is somewhat counter-intuitive: If $\theta_0=0$, the $+1$ defect has pure \emph{splay} near the core, and its velocity is especially sensitive to the Frank constant for \emph{bend}. If $\theta_0=\pi/2$, the $+1$ defect has pure \emph{bend} near the core, and its velocity is especially sensitive to the Frank constant for \emph{splay}.

This counter-intuitive dependence can be understood through the following argument: When a splay defect with $\theta_0=0$ in Fig.~1(a) moves to the right, the white splay area remains approximately unchanged, but the black bend area between the test defect and the boundary defect shrinks. The dynamics is controlled by the energy cost of the shrinking bend area, which is given by $k_B$. Conversely, when a bend defect with $\theta_0=\pi/2$ in Fig.~1(b) moves to the right, the black bend area remains approximately unchanged, but the white splay area between the test defect and the boundary defect shrinks. The dynamics is controlled by the energy cost of the shrinking splay area, which is given by $k_S$.

\begin{figure}
\includegraphics[width=\columnwidth]{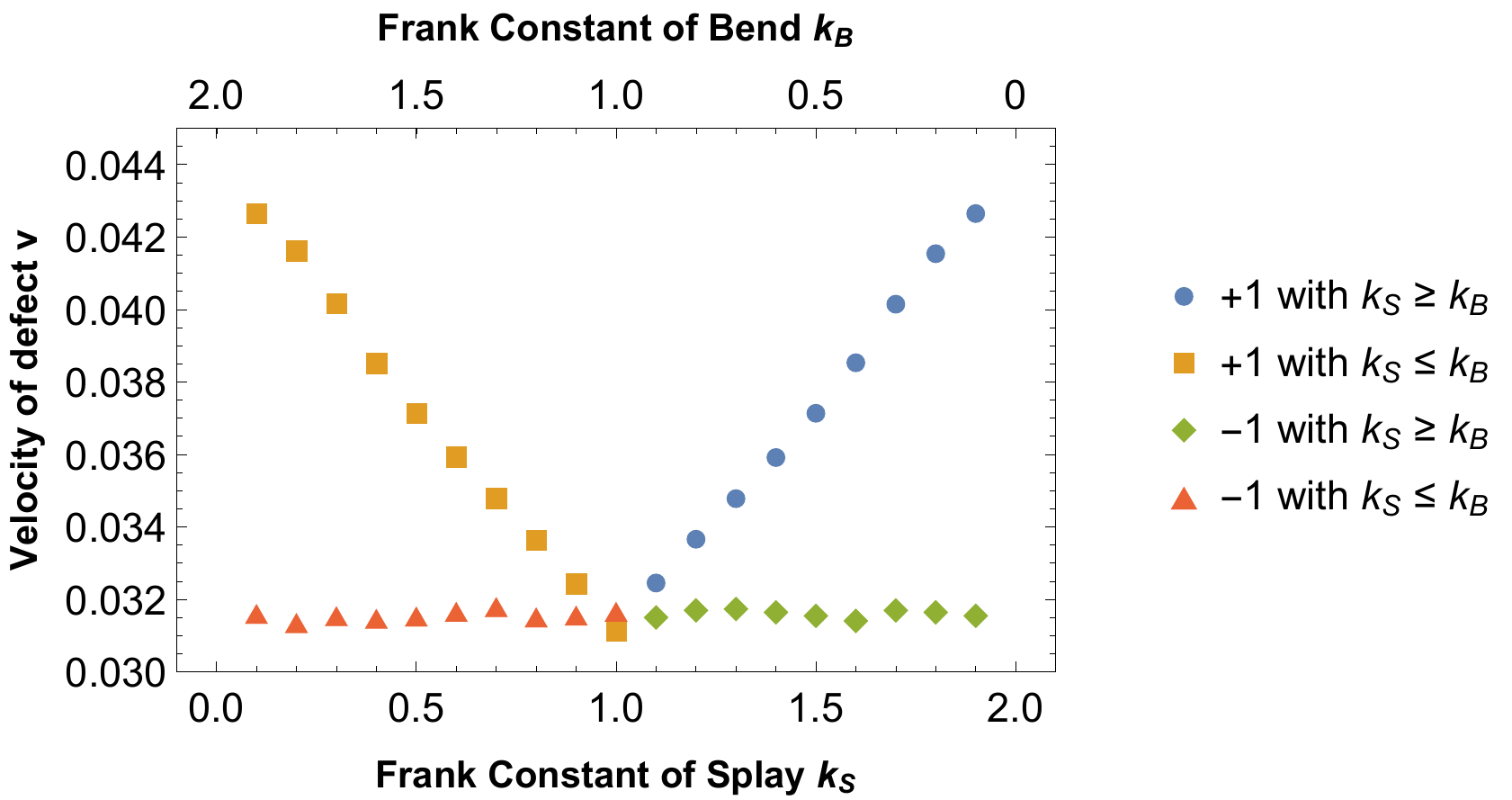}
\caption{Velocity of defects with different topological charges when one Frank constant increases while the other decreases, so that the sum is constant. When $k_S<k_B$, we simulate a $+1$ test defect or boundary defect with pure splay near the core. When $k_S>k_B$, we simulate a $+1$ test defect or boundary defect with pure bend near the core.}
\end{figure}

For another view of the effects of Frank constants on defect velocity, we simulate the same geometry but now increase one Frank constant while decreasing the other, so that the total remains constant. In Fig.~3, $k_S$ increases from 0.1 to 1.9, and $k_B$ decreases from 1.9 to 0.1, with the constant sum $k_S+k_B=2$. When $k_S<k_B$, the $+1$ defect has pure splay near its core, and its velocity is most sensitive to the bend elastic constant. As a result, the velocity increases when $k_B$ increases and $k_S$ decreases. Similarly, $k_B<k_S$, the $+1$ defect has pure bend near its core, and its velocity is most sensitive to the splay elastic constant. Its velocity increases when $k_S$ increases and $k_B$ decreases. By contrast, for a $-1$ defect, the velocity remains constant when one Frank constant increases while the other decreases, because the average Frank constant is constant.

From these results, we see that unequal Frank constants influence a $+1$ defect much more than a $-1$ defect. For a $+1$ defect, the angle $\theta_0$ determines whether a defect has pure splay or pure bend near the core. When the Frank constants are unequal, this angle is driven toward the value corresponding to the lower-energy mode. The $+1$ defect velocity is then controlled mainly by the Frank constant corresponding to the higher-energy mode. For a $-1$ defect, the angle $\theta_0$ determines only the orientation of the defect, not the defect structure. This orientation is free to vary even when the Frank constants are unequal. The $-1$ defect velocity is controlled by the average of the Frank constants.

\section{Trajectory of defect pair annihilation}

In this section, we investigate the annihilation process for a pair of $\pm1$ defects. We emphasize that the initial state of the two defects may have the optimal relative orientation, or it may have some non-optimal relative orientation. In our previous work~\cite{Tang2017}, we used conformal mapping to calculate the lowest-energy director configuration around any two defects, with the assumption of \emph{equal} Frank constants. For arbitrary defect positions $\mathbf{R}_1=(x_1,y_1)$ and $\mathbf{R}_2=(x_2,y_2)$, arbitrary topological charges $k_1$ and $k_2$, and arbitrary relative orientation, the result is
\begin{align}
\label{arbitraryorientation}
\theta(\mathbf{r})={}&k_1\tan^{-1}\left[\frac{y-y_1}{x-x_1}\right]+k_2\tan^{-1}\left[\frac{y-y_2}{x-x_2}\right]\\
&+\frac{\delta\theta}{2}
\left[1+\frac{\log(|\mathbf{r}-\mathbf{R}_1 |^2)-\log(|\mathbf{r}-\mathbf{R}_2 |^2)}{\log(|\mathbf{R}_1 -\mathbf{R}_2 |^2)-\log(r_\text{core}^2)}\right]
+\Theta,\nonumber
\end{align}
where 
\begin{align}
&\delta\theta = \theta_2 - \theta_1 + k_2 \tan^{-1}\left[\frac{y_1-y_2}{x_1-x_2}\right] - k_1 \tan^{-1}\left[\frac{y_2-y_1}{x_2-x_1}\right],\nonumber\\
&\Theta = \theta_1 - k_2 \tan^{-1}\left[\frac{y_1-y_2}{x_1-x_2}\right].
\end{align}
The resulting elastic free energy is
\begin{align}
F={}&\pi K (k_1 + k_2)^2 \log\left[\frac{R_\text{max}}{r_\text{core}}\right] 
- 2\pi K k_1 k_2 \log\left[\frac{|\mathbf{R}_1 -\mathbf{R}_2 |}{2r_\text{core}}\right]\nonumber\\
&+ \frac{\pi K \delta\theta^2}{2} \frac{\log[|\mathbf{R}_1 -\mathbf{R}_2 |/(2r_\text{core})]}{[\log(|\mathbf{R}_1 -\mathbf{R}_2 |/r_\text{core})]^2} .
\label{interaction}
\end{align}
Here, the first term is the free energy cost associated with the total topological charge $(k_1 + k_2)$ in a system of size $R_\text{max}$; it vanishes for a pair of oppositely charged defects with $k_1=-k_2$. The second term is the interaction free energy of two defects with the optimal relative orientation, which varies logarithmically with the separation. The third term is the extra interaction free energy associated with a non-optimal relative orientation. The optimum occurs at $\delta\theta=0$.

In the special case of optimum relative orientation $\delta\theta=0$, the interaction of Eq.~(\ref{interaction}) is a central-force interaction, which depends on the defect positions $\mathbf{R}_1$ and $\mathbf{R}_2$ only through the separation ${|\mathbf{R}_1 -\mathbf{R}_2 |}$. In this special case, the interaction between opposite topological charges attracts the defects straight together. However, in the general case of $\delta\theta\not=0$, this interaction is \emph{not} a central-force interaction, because it depends on $\delta\theta$, and $\delta\theta$ depends on the orientation of the vector ${\mathbf{R}_1 -\mathbf{R}_2}$. Hence, in that general case, the interaction does not only attract the defects together; it also causes them to move transversely to the line between them, in order to reduce the $\delta\theta^2$ term in the interaction. This combination of effects leads to an S-shaped annihilation trajectory. Indeed, numerical simulations have found S-shaped trajectories of defect annihilation, in models with equal Frank constants~\cite{Vromans2016,Tang2017}.

At this point, the question is: What is the defect annihilation trajectory if the Frank constants are \emph{unequal?} To address this question, we set up an initial condition with two defects, with the director configuration of Eq.~(\ref{arbitraryorientation}), and then allow this configuration to relax following the dynamics of Eq.~(\ref{dynamicequations}).

\begin{figure}
\includegraphics[width=\columnwidth]{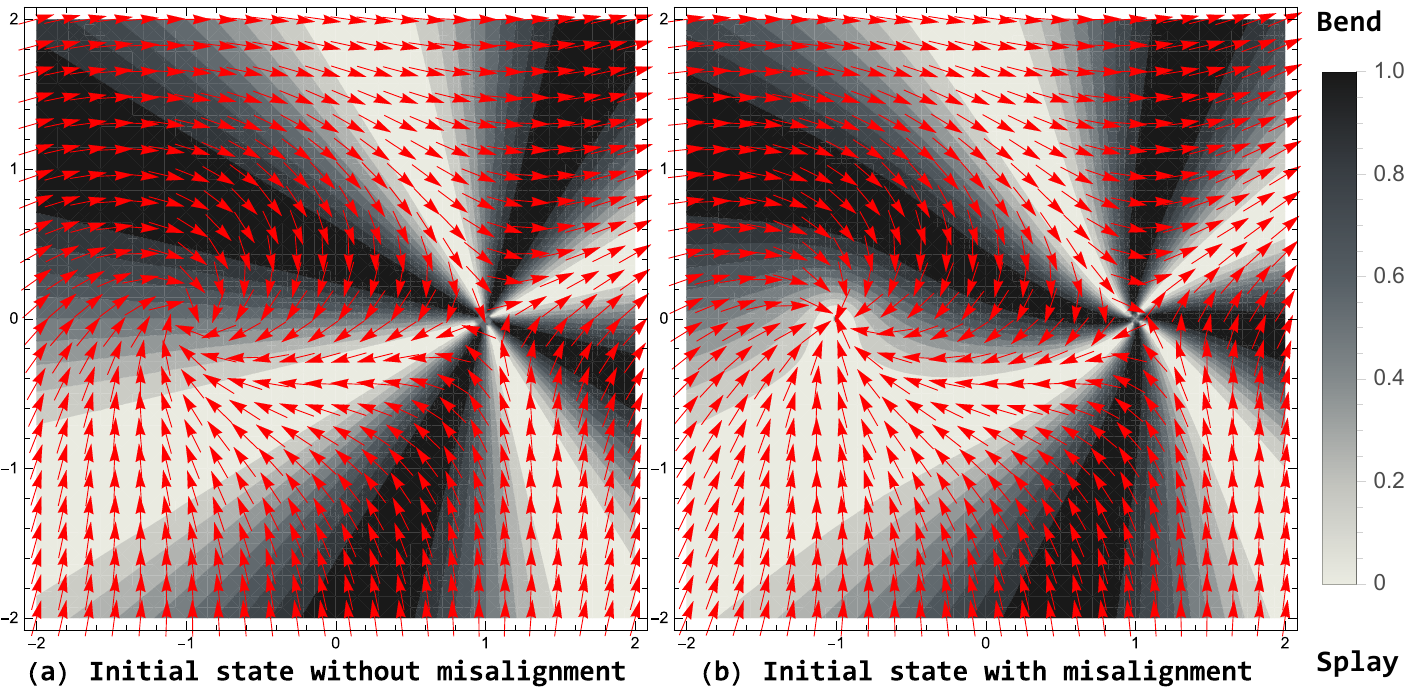}
\caption{Initial conditions for defect annihilation simulations. (a)~Optimal relative orientation with $\delta\theta=0$ and $\Theta=\pi/4$. (b)~Misalignment with $\delta\theta=\pi/2$ and $\Theta=0$. In both cases, the splay area (white) is below the $+1$ defect, and the bend area (black) is above it.}
\end{figure}

Figure 4 shows two examples of the initial condition, with a $k_1=+1$ defect at $\mathbf{R}_1=(-1,0)$ and a $k_2=-1$ defect at $\mathbf{R}_2=(1,0)$. In Fig.~4(a) the two defects have the optimal relative orientation $\delta\theta=0$, and in Fig.~4(b) they have a misalignment of $\delta\theta=\pi/2$. The background color indicates the regions of mostly splay in white, and mostly bend in black, calculated with equal Frank constants. In each case, the overall constant $\Theta$ is chosen so that the splay region is below the $+1$ defect, and the bend region is above it.

\begin{figure}
\includegraphics[width=\columnwidth]{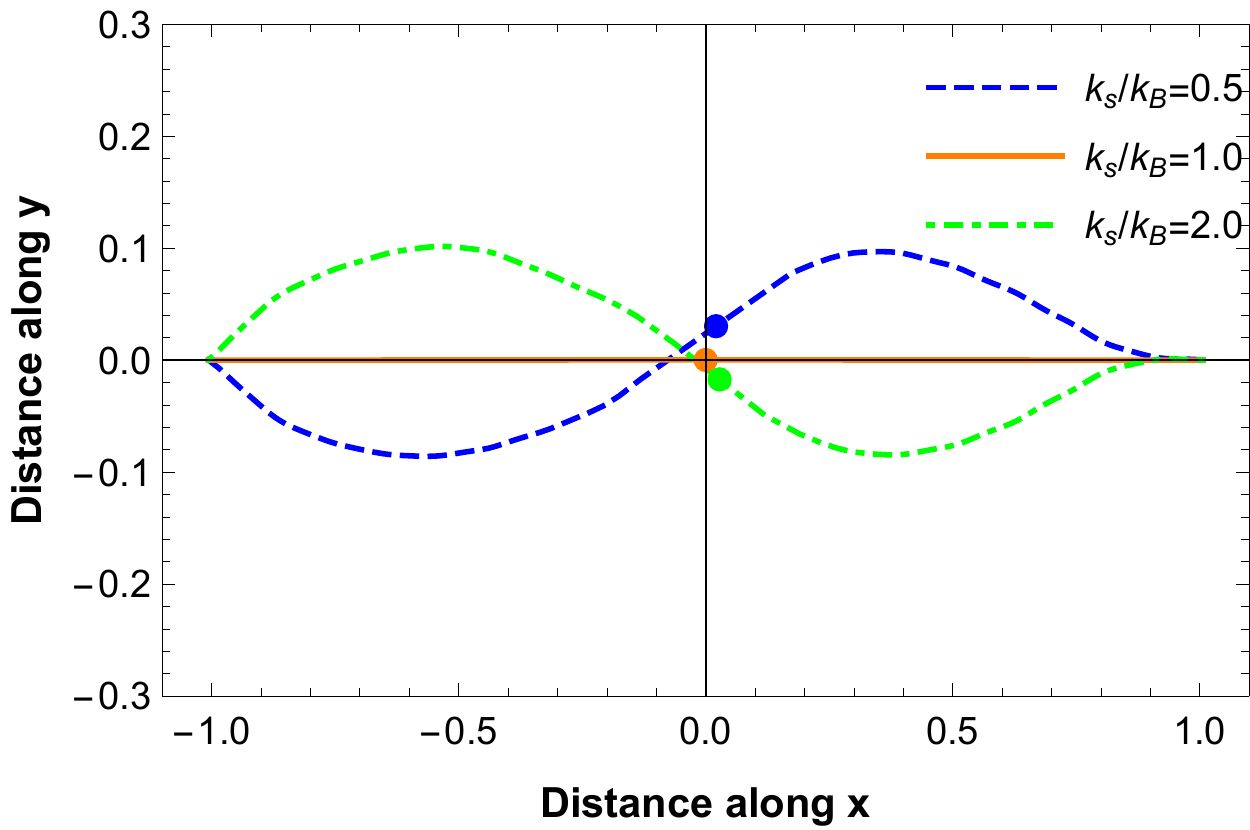}
\caption{Defect annihilation trajectories for different ratios of Frank constants, with no defect misalignment in the initial conditions.}
\end{figure}

In Fig.~5, we see the time evolution of the structure of Fig.~4(a), with no initial misalignment. The solid orange line shows the behavior when $k_S/k_B=1.0$. In this case, the defects move straight together, as expected from the central-force attraction of Eq.~(\ref{interaction}). The annihilation point is exactly in the middle, showing that the defects move with equal speeds, as expected for a model with no backflow.

By comparison, the green dotted dashed line shows the evolution when $k_S/k_B=2.0$. Here, the defects move in an S-shaped trajectory only because of the unequal Frank constants, even without initial misalignment. This trajectory can be understood through the following argument: Early in the time evolution, the $+1$ defect is driven to pure bend near its core because $k_B<k_S$. This bend defect moves upward into the mostly bend region to minimize the Frank elastic energy. The $-1$ defect compensates by moving downward, but this compensation occurs somewhat later, as can be seen by the early parabolic trajectory of the $-1$ compared with the early linear trajectory of the $+1$. This combination of defect movements makes the S-shape. The annihilation point is slightly closer to the initial position of the $-1$ than the $+1$, indicating that the $+1$ moves slightly more rapidly than the $-1$. This difference of speed occurs purely because of the unequal Frank constants, even without backflow.

The blue dashed line shows the corresponding evolution when $k_S/k_B=0.5$. This behavior is analogous to the previous case, but with splay and bend reversed. Again, the defects move in an S-shaped trajectory even without initial misalignment, and the $+1$ has a higher speed than the $-1$ even without backflow. Both of these effects occur only because of the unequal Frank constants.

\begin{figure}
\includegraphics[width=\columnwidth]{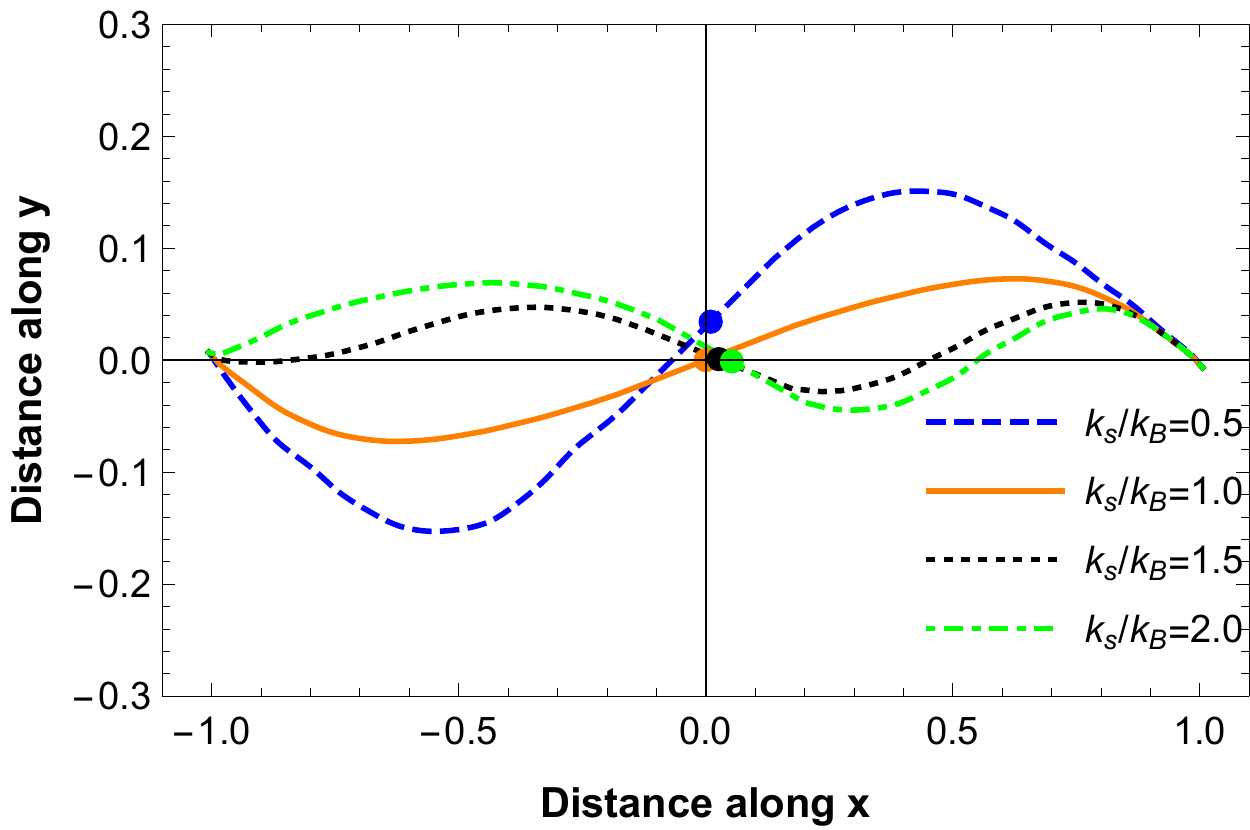}
\caption{Defect annihilation trajectories for different ratios of Frank constants, with an initial defect misalignment of $\delta\theta=\pi/2$.}
\end{figure}

In Fig.~6, we see the corresponding results for the time evolution of the structure of Fig.~4(b), with an initial defect misalignment $\delta\theta=\pi/2$. The solid orange line shows the behavior when $k_S/k_B=1.0$. The defects move in an S-shaped trajectory because of the initial misalignment, even with equal Frank constants. The annihilation point is exactly in the middle, showing that the defects move with equal speeds. The blue dashed line shows the evolution when $k_S/k_B=0.5$. Here, the S-shape is enhanced, because the effects of the defect misalignment and the unequal Frank constants add together. The black dotted line shows the evolution when $k_S/k_B=1.5$. In this case, the S-shape is reduced, because the effects of the defect misalignment and the unequal Frank constants approximately cancel each other. Finally, the green dotted dashed line shows the ratio $k_S/k_B=2.0$, where the effect of unequal Frank constants dominates the defect misalignment.

\section{Discussion}

In this article, we have investigated the dynamics of defect motion in a 2D smectic-C liquid crystal film, and we have found that the the annihilation trajectory is influenced by the difference of Frank constants. This effect arises mainly because of the special features of the $+1$ defect: Any difference of Frank constants drives this defect into a state of pure splay or pure bend near the core, and the pure splay or bend changes the motion of the defect. In particular, a pure splay or bend defect tends to move into a region with the corresponding deformation, and its speed depends mainly on one Frank constant. By contrast, a $-1$ defect necessarily includes both splay and bend deformations, and its speed depends on the average of the Frank constants. Hence, a difference of Frank constants causes the defects to move in an S-shaped annihilation trajectory, with the $+1$ defect moving more rapidly than the $-1$ defect.

In liquid crystal research, the S-shaped trajectory and the difference in defect speeds are generally attributed to other causes. The S-shaped trajectory is normally associated with initial defect misalignment, and the difference in defect speeds is normally associated with backflow. We now see that the difference of Frank constants provides an alternative mechanism for both of these effects. In simulations, it is simple to distinguish these mechanisms by varying the Frank constants to be equal or unequal. In experiments, however, these mechanisms occur simultaneously; an experimental liquid crystal has unequal Frank constants as well as defect misalignment and backflow. Hence, this study provides a caution for identifying the mechanism for experimental defect motion.

\acknowledgments

We would like to thank R. Stannarius and K. Harth for helpful discussions. This work was supported by National Science Foundation Grant DMR-1409658.

\bibliography{FrankConstants2}

\end{document}